\documentclass[article,twoside]{revtex4}
\usepackage{color}
 \usepackage{amsbsy}
  \usepackage{bm}
 \usepackage{amsfonts}
\usepackage{amssymb}
\usepackage{amsmath}
\usepackage{graphics}
\usepackage{graphicx}
\begin{document}

\title{Coexisting phases in the chiral transition within the Linear sigma model with quarks}

\author{R. M. Aguirre}
\affiliation{Departamento de Matematica, Universidad Nacional de
La Plata\\ and Instituto de Fisica La Plata,  CONICET\\ Argentina}

\begin{abstract}
It is believed at present that the chiral transition changes from
a smooth crossover to a first-order transition at low temperatures
and high densities. Such regime is commonly analyzed using
effective models since first principle calculations, as in lattice
arrangements, are not feasible. This transition is assumed to be
discontinuous, with unstable or metastable intermediate states.
However, if multiple charges are simultaneously conserved the
system could undergo a continuous change through a coexistence of
equilibrium states. This type of transition has multiple
manifestations, as in the nuclear liquid-gas transition causing
the spinodal fragmentation. The coexistence of phases in the
chiral transition is studied here for quark matter assuming the
conservation of the isospin composition.  Using the Linear sigma
model with quarks several remarkable effects are found and
discussed.\end{abstract}

\maketitle

\section{INTRODUCTION}
The study of matter under extreme conditions of density and
temperature has been intensively developed in the last decades
with the aim of scrutinizing the different regimes of the strong
interaction. Experiments on heavy ion collisions have provided a
lot of evidence to understand the high temperature and low matter
concentration domain. Complementary information has been obtained
from the observational data of compact stars, which corresponds to
low temperature and medium-high densities. Some additional
physical precisions have been obtained from the lattice simulation
of the fundamental theory (LQCD) \cite{ALFORDK,SON,NISHIDA,EJIRI}.
This method is particularly appropriate to study the vacuum
properties and finite temperature effects, although finite density
systems are not accessible due to the well known problem of the
negative sign. To avoid such failure, different approaches have
been proposed, as for instance the use of an imaginary baryonic
chemical potential $\mu_B$, or the introduction of an isospin
chemical potential $\mu_I$ \cite{ALFORDK}, which takes into
account  the conservation of the flavor degree of freedom.
Furthermore, theoretical investigations using effective models
have tried to compile the phenomenology within a unified
description. As an outcome of these common efforts, a picture of a
complex phase diagram has emerged, with several phase transitions,
collective states, etc.
\\
Special attention was paid to the deconfinement phase transition
\cite{ARYAL,PRAKASH} since its realization could have clear
manifestations in heavy ion collision events or, as pointed out
more recently, it could leave its imprint in the gravitational
waves  coming from the collapse of companion neutron stars.
\\
The fluctuations of conserved charges, such as baryon number or
electric charge, have been proposed as an efficient signal of the
transition and consequently have been focused by the LQCD . They
have been related to the corresponding thermodynamical
susceptibilities by the fluctuation dissipation theorem and have
been used to reconstruct the equation of state.

The chiral phase transition has also been an object of theoretical
investigation. As in the case of the deconfinement, the general
consensus is that its character changes from first order at low
temperatures to second order at higher ones. In the case that non
zero quark mass is considered, the second order transition becomes
a soft crossover. As a consequence, the line of transition points
in the $(\mu_B,T)$ plane exhibits a critical endpoint (CEP)
separating both regimes.\\

Systems at zero baryon density ($\mu_B=0$) but with an isospin
imbalance ($\mu_I\neq 0$) have been hypothesized to be in a pion
condensed phase before the chiral symmetry is restored and $\mu_I
> m_\pi$ \cite{SON}. This is an interesting assumption that relies
on the identification of $\mu_I$ as a gauge field. However, it was
appropriately warned that a system with $\mu_I > \mu_B$ will be
physically unstable by weak decay \cite{SON}.
\\
General features of the chiral transition in quark matter with
isospin imbalance have been investigated in
\cite{TOUBLAN,FRANK,BARDUCCI,BARDUCCI1,LOEWE,HE,HE1,ZHANG,MUKHERJEE,KOVACS0,MU,ABUKI,XIA,STIELE,LIU,AVANCINI,LOPES,ADHIKARI,COSTA,CHU,AYALA,AYALA0}.
The coherence of the theoretical predictions with the LQCD results
at zero baryon number has been particularly analyzed in
\cite{LOEWE,AVANCINI,LOPES,ADHIKARI,AYALA0}.\\
In this paper, instead, quark matter at finite baryon density and
with isospin asymmetry is considered. For this purpose an
effective model, the Linear Sigma Model with quarks (LSMq), is
adopted and the weak interaction is neglected as in \cite{SON}.
Thus, the $u$ and $d$ quark flavors can be regarded as stable
degrees of freedom, and an additional meson  $\zeta$ with isospin
structure is introduced and treated on the same foot as the
standard $\sigma, \pi$ ones. Several attempts to enlarge the meson
family within the LSMq have been made
\cite{PARGANLIJA,PARGANLIJA1,KOVACS}, particularly the role of the
scalar, isovector meson has been discussed from long time ago
\cite{GASIOROWICZ,METZGER,LENAGHAN,SCHAEFER,ABUKI,AYALA1}. In
order to stress the results on thermodynamical instabilities, the
simplicity of the theoretical formulation is preferred here.
Hence, only the meson fields $\sigma,\, \pi, \,\zeta$ are
considered, and a more sophisticated approach is left for further
development.

The remaining of this work is organized as follows, the mean field
approach to the LSMq, also known as Quark Meson model, is
presented in the next section. The results for the thermodynamics
of the system is discussed in Sec. \ref{RESULTS}, and the final
conclusions are drawn in Sec. \ref{SUMMARY}.
\section{THE MODEL}
The LSMq is an effective model of the strong interaction, adequate
for the  low energy regime. It has been used  extensively due to
its versatility and simplicity. The concise Lagrangian density
used in this work is \cite{AYALA1}
\begin{eqnarray}{\cal
L}&=&\,\bar{\Psi}\left( i \not \!
\partial+g \,\Phi \right)\Psi+\frac{1}{2}\left(\partial_\mu \sigma
\partial^\mu \sigma+ \partial_\mu \bm{\pi} \cdot
\partial^\mu \bm{\pi}+\partial_\mu \bm{\zeta} \cdot
\partial^\mu \bm{\zeta}\right)+\frac{1}{4}C_0 \text{Tr}
\left(\Phi^\dag \Phi\right)\nonumber \\
&&-\frac{1}{16} C_3 \left[\text{Tr} \left(\Phi^\dag
\Phi\right)\right]^2-\frac{1}{8} C_2 \text{Tr} \left(\Phi^\dag
\Phi\right)^2-h\, \sigma \label{LAGRANGE}
\end{eqnarray}
with the quark bi-spinor $\Psi=\left(\psi_u\;\psi_d\right)^t$, the
iso-multiplets scalar $\bm{\zeta}$, and pseudo-scalar $\bm{\pi}$,
and $\Phi=\sigma+i\,\gamma_5 \bm{\pi}\cdot
\bm{\tau}+\bm{\zeta}\cdot \bm{\tau}$. Flavor degeneracy in the
quark masses  is assumed and a symmetry breaking term, linear in
$\sigma$  has been included.

The scalar meson fields can be decomposed as the sum of a thermal
expectation value and its fluctuation $\sigma=s+\delta \sigma$,
$\zeta_a=z\,\delta_{a3}+\delta \zeta_a$. Neglecting the
fluctuations leads one to the mean field approach ($MFA$), while
its explicit consideration can be arranged as higher order
corrections \cite{DOLAN,PETROPOULOS,CORNWALL}. Here I adopt the one-loop
approach, as in \cite{AYALA0}, where mesons behave as free
particles with effective masses. Hence the grand potential per
unit volume of uniform quark matter can be written as
\begin{eqnarray}
\omega(T,\mu)&=&\omega_{\text{vac}}-\frac{N_c}{\beta
\pi^2}\sum_{j=u,d}\int_0^\infty
dp\,p^2\left[\text{log}\left(1+e^{-\beta(E_p-\mu_j)}\right)
+\text{log}\left(1+e^{-\beta(E_p+\mu_j)}\right)\right] \nonumber
\\&&+\frac{1}{2 \beta
\pi^2}\sum_{\alpha=\sigma,\pi_a,\zeta_a}\int_0^\infty
dp\,p^2\,\text{log}\left(1-e^{-\beta\,E_\alpha}\right)
\label{GRAND P}
\end{eqnarray}
The vacuum term is divergent, but a finite contribution can be
extracted by dimensional regularization followed by an appropriate
subtraction scheme
\begin{equation}\omega_{\text{vac}}=\frac{N_c}{8
\pi^2}\sum_{j=u,d}\left[m_j^4 \,\text{log}\left(\frac{m_j}{m_0}
\right)+\frac{m_0^4-m_j^4}{4}\right] -\frac{1}{32 \pi^2}
\sum_{\alpha=\sigma,\pi_a,\zeta_a} \left[M_\alpha^4
\,\text{log}\left(\frac{M_\alpha}{M_{0 \alpha}} \right)+\frac{M_{0
\alpha}^4-M_\alpha^4}{4}\right] \label{GP VAC}\end{equation}

The notation $m_j=g(s+I_j \, z)$ has been introduced for the
effective quark masses, where $I_j=1\, (-1)$ for $j=u\, (d)$. The
subindex $0$ in the masses indicates the physical values.

Within the formalism of \cite{CORNWALL,CAMELIA}, the effective
potential is a functional of the mean values $s,\, z$ and of the
full propagators of the mesons. Hence the definitions of $s,\, z$
are given,as usual, by the extremum conditions
\[ 0=\frac{\partial \omega}{\partial s},\;\;\;\;\;\; 0=\frac{\partial \omega}{\partial z}\]
Giving
\begin{equation}
0=(-C_0+C_3 s^2+C_4 z^2)\,s-h-\frac{g
N_c}{\pi^2}\sum_{j=u,d}m_j\left[\frac{m_j^2}{2}\,\text{log}\left(\frac{m_j}{m_0}
\right)-\int_0^\infty \frac{dp\; p^2}{E_j}\left(n_{F j}+\bar{n}_{F
j}\right) \right],\label{S MFA}
\end{equation}
\begin{equation}
0=(-C_0+C_3 z^2+C_4 s^2)\,z-\frac{g N_c}{\pi^2}\sum_{j=u,d}I_j
\,m_j\left[\frac{m_j^2}{2}\,\text{log}\left(\frac{m_j}{m_0}
\right)-\int_0^\infty \frac{dp\; p^2}{E_j}\left(n_{F j}+\bar{n}_{F
j}\right) \right],\label{Z MFA}
\end{equation}
respectively.  In addition, there is a condition for each of the
mesons, which in the present approach reduces to
$M_\sigma^2=-C_0+3 C_3 \,s^2+(C_3+2 C_2) \,z^2$,
$M_{\pi\,a}^2=-C_0+C_3 (s^2+z^2)$, $M_{\zeta\,a}^2=-C_0+C_3\,
n_a\,z^2+(C_3+2 C_2) \,s^2$ and $n_a=1\, (3)$ for $a=1,2\,(3)$.\\
In Eqs.(\ref{S MFA}, \ref{Z MFA}) the equilibrium Fermi
distribution functions $n_{F j}(T,\mu_j)$ for quarks and
$\bar{n}_{F j}(T,\mu_j)$ for antiquarks have been introduced.

The entropy of the system is given by the thermodynamical
definition ${\cal S}=S/V=-\partial\,\omega/\partial T$ as
\begin{eqnarray}
{\cal
S}(T,\mu)&=&-\frac{\omega-\omega_{\text{vac}}}{T}+\frac{N_c}{
\pi^2 T}\sum_{j=u,d}\int_0^\infty dp\,p^2\left[(E_j-\mu_j)\, n_{F
j}+(E_j+\mu_j)\, \bar{n}_{F j}\right] \nonumber
\\&&+\frac{1}{2 \pi^2 T}\sum_{\alpha=\sigma,\pi_a,\zeta_a}
\int_0^\infty dp\,p^2\,E_\alpha\,n_{B \alpha}.\label{ENTROPY}
\end{eqnarray}
In the last equation the equilibrium Bose distribution function
$n_B(T)$ is used. Finally, the energy density is evaluated by the
Legendre transform ${\cal E}=\omega+ T \,{\cal
S}+\sum_{j=u,d}\mu_j\,n_j$, using the quark number density
\[n_j=\frac{N_c}{\pi^2}\int_0^\infty dp\,p^2\left( n_{F
j}- \bar{n}_{F j}\right).\]

In the following the evolution of matter at fixed isospin fraction
$x=(n_d-n_u)/(n_u+n_d)$ is considered. Some characteristic
quantities  of the thermodynamical evolution of the system are the
isothermal speed of sound $v=-\left(\partial \omega/\partial {\cal
E}\right)_T$ and the second order susceptibilities $\chi$
\cite{GAVAI0,FERRONI,RATTI,SASAKI,SASAKI1,SASAKI2,STOKIC,SKOKOV,ALMASI,SKOKOV,SKOKOV1,SKOKOV2}.
The former has been demonstrated to be useful to indicate the
emergence of new degrees of freedom. Meanwhile the
susceptibilities manifest the dynamical fluctuations
characterizing the bulk (semi-classical) behavior of the system
\cite{ASAKAWA,GAVAI,DATTA}. Fluctuations are closely related to
phase transitions, in particular those related to conserved
charges. For such reason the susceptibilities associated with
conserved charges have been
focused within the LQCD \cite{RATTI,GAVAI,DATTA}. \\
In the following the susceptibilities associated with the quark numbers
\[ \chi_B=\left(\frac{\partial n_B}{\partial \mu_B}\right)_{\mu_3},\;\;
\chi_3=\left(\frac{\partial n_3}{\partial \mu_3}\right)_{\mu_B},\]
will be considered. Here the densities for the baryon number
$n_B=(n_d+n_u)/3$, and the isospin number $n_3=n_d-n_u$ have been
introduced, together with the chemical potentials for the baryon
$\mu_B=3(\mu_d+\mu_u)/2$ and the isospin $\mu_3=\mu_d-\mu_u$
charges.

\section{RESULTS AND DISCUSSION}\label{RESULTS}

The parameters shown in Eq. (\ref{LAGRANGE}) depend on the
physical masses of the mesons and the pion decay constant as
\[C_0=\frac{M_{\sigma 0}^2-3 M_{\pi 0}^2}{2},\; C_2=\frac{M_{\zeta 0}^2-3 M_{\pi 0}^2}{2 f_\pi^2},
C_3=\frac{M_{\sigma 0}^2- M_{\pi 0}^2}{2 f_\pi^2},\]
and the values $M_{\sigma 0}=500$ MeV, $M_{\pi 0}=138$ MeV, and
$M_{\zeta 0}=984$ MeV have been used. In particular the mass of
the $\zeta$ meson has been identified with that of the $a_0$(980)
since this is the lightest manifestation in the hadronic sector of
an scalar isoscalar. The coupling constant $g$ is related to the
quark constituent mass through $m_{q 0}=g\,f_\pi$ and will be
defined in the next step.
\\
The model is used in the following to analyze the chiral phase
transition in homogeneous quark matter. The relation between
chiral breakdown and deconfinement is not clear yet, and for the
sake  of simplicity it is not treated here. However, this subject
has been extensively studied, specially with the aid of a
phenomenological Polyakov potential. Under the conditions
considered in this work the isospin chemical potential verifies
$\mu_3<M_{\pi 0}$, for this reason the pion condensation is not
taken into account.
\\
As finite density matter is considered the approach turns
inadequate since the pion mass becomes imaginary. As for example,
in flavor symmetric matter it is found that $M_\pi^2<0$ as soon as
$s/f_{\pi}<0.9$. In consequence the contributions of all the
mesons, as in  Eqs.(\ref{GRAND P}), (\ref{GP VAC}),
(\ref{ENTROPY}), are disregarded in the following but keeping the
one-loop approach for quarks. A similar treatment is usual in
relativistic hadronic models, and it is commonly applied to study
phase transitions as the nuclear liquid-gas, or deconfinement in
high density matter. The MFA is also the standard approach in
quark matter calculations using the effective Nambu-Jona Lasinio
model \cite{SASAKI,SASAKI1,SASAKI2}, and it has been used in the
case of the LSMq model too \cite{STOKIC}.

\begin{figure}
    \centering
    \includegraphics[width=0.8\textwidth]{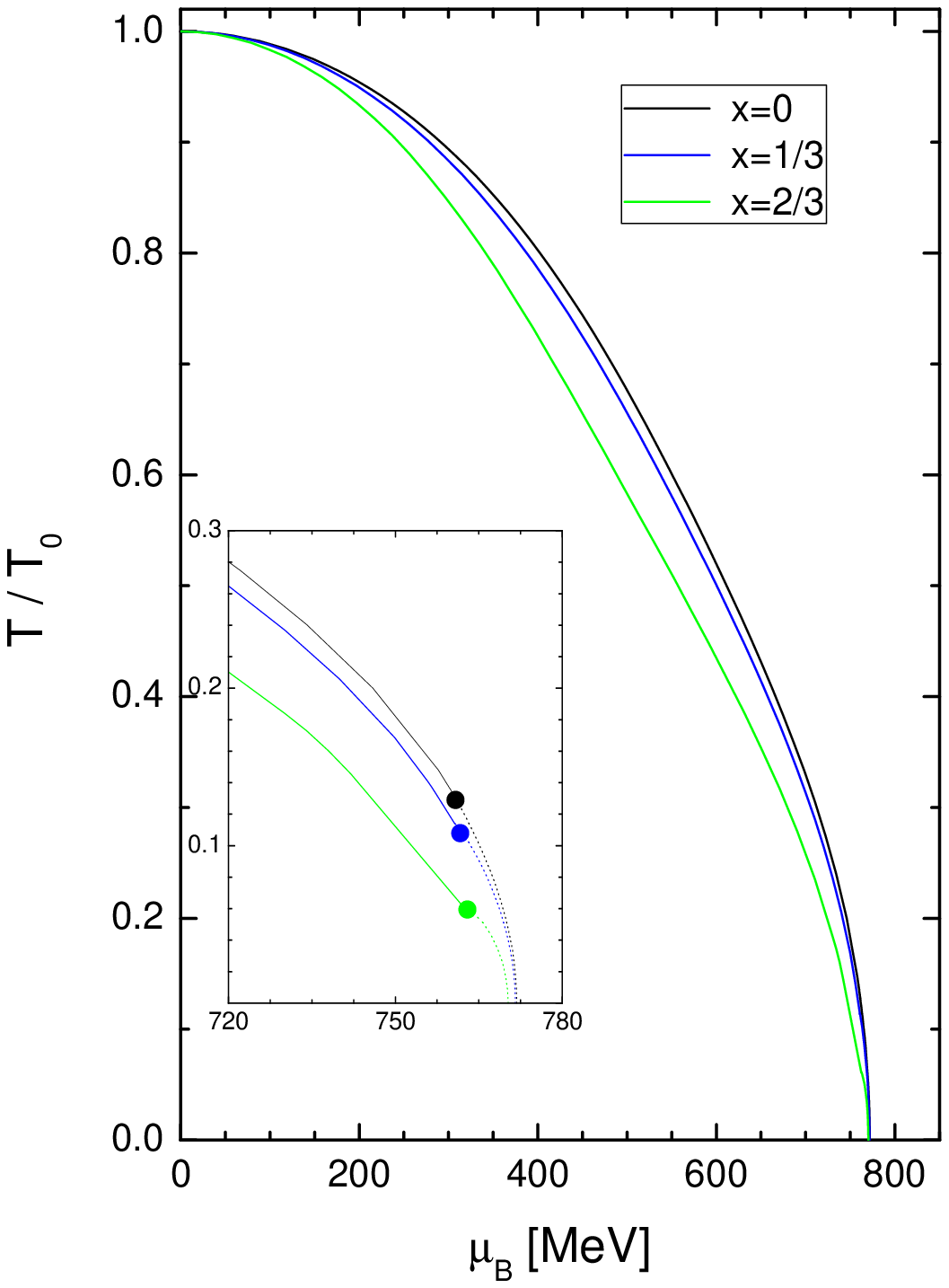}
    \caption{The transition temperature in terms of the bayon number
    chemical potential for several global isospin parameters $x$. The insertion
    shows details of the first order transition (dashed lines), the CEPs (full circles)
    and the crossover transition (full lines). Within the ECR the Eq.(\ref{x Binodal}) applies.  }
    \label{Fig Critical}%
\end{figure}

\begin{figure}
    \centering
    \includegraphics[width=0.8\textwidth]{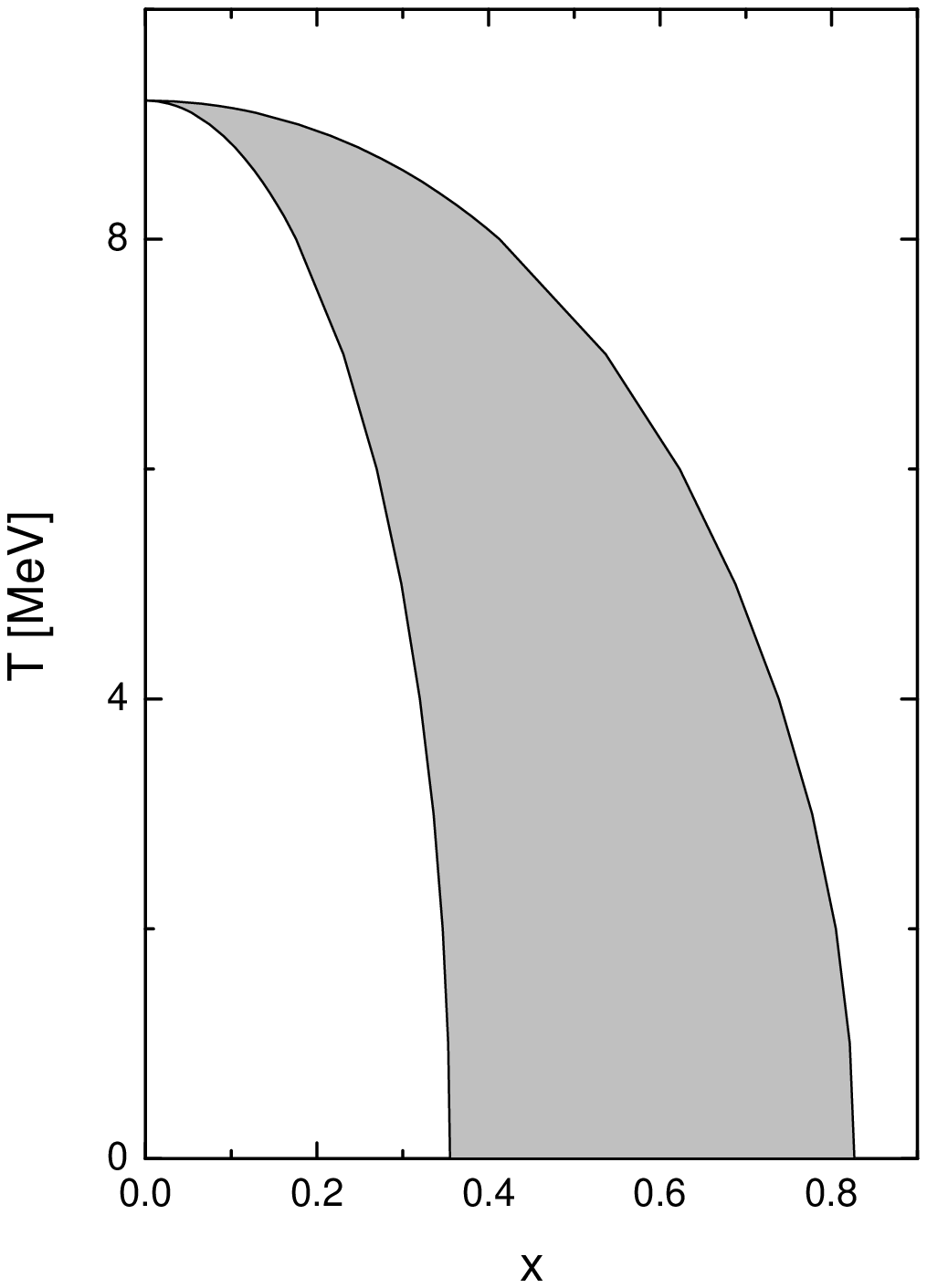}
    \caption{An isobar section of the equilibrium coexistence region in the $x-T$ plane for  $P=45$ MeV fm$^{-3}$.
    }
    \label{Fig Binodal}%
\end{figure}

 At present it is believed that the chiral transition is a
continuous crossover for low densities and becomes first order for
sufficiently high values of $\mu_B$. There is no clear definition
of the transition temperature in a crossover, but to show a
characteristic temperature we adopt the value for the inflection
point of the chiral order parameter \cite{PAWLOWSKI}, i.e. where
the minimum of $s^2+z^2$ occurs. The critical temperature  $T_0$
is  found to be a decreasing function of $g$ and low values of
$m_{q 0}$ favors the compatibility with LQCD estimations
\cite{EJIRI}. However, in the present approach it is found that
for the lower $m_{q 0}$ the expected change of regime for high
$\mu_B$ is lost. Therefore an intermediate value is adopted
$m_{q0}=250$ MeV, which yields the phase diagram shown in
Fig.\ref{Fig Critical} for several values of $x$. The value
obtained $T_0 \simeq 120$ MeV is out the range $160-180$ MeV
predicted by LQCD, but a better fit could be obtained by including
higher order corrections to the present scheme
\cite{PETROPOULOS,AYALA1}. In regard of the values selected for
the flavor asymmetry,  $x=0$ corresponds to the flavor symmetric
case usually discussed in finite density calculations, $x=1/3$
represents electric charge neutral matter similar to that found in
self-bound quark stars, and finally $x=2/3$ is an extrapolation to
the negatively charged case.\\
As expected a change of regime is found with a CEP
in between. For $T=0$ and $\mu_B=0$ all the curves coincide, while
at intermediate values of the chemical potential the
pseudo-critical
temperature decreases with $x$.\\
The first order phase transition is generated by instabilities in
the equation of state, since the thermodynamic potential $\omega$
does not vary monotonically with the chemical potentials.
Depending on the value of the surface tension, the transition
occurs as a discontinuous change or as a continuous passage with
an intermediate coexistence region. The last case is feasible
because there are multiple charges conserved simultaneously
\cite{GLENDENNING,HEISELBERG}, and both coexisting phases
correspond to the regime of broken chiral symmetry. Assuming a low
surface tension the system evolves by following the Gibbs
construction, and a parameter $\lambda$ is introduced which
indicates the relative abundance of the coexisting phases, named
$a$ and $b$ in the following. Hence the equilibrium conditions
read
\begin{equation}T_a=T_b,\;\; \mu_B^a=\mu_B^b,\;\;
\mu_3^a=\mu_3^b,\label{GIBBS1}\end{equation}
\begin{equation}P_a(T_a,\mu_B^a,\mu_3^a)=P_b(T_b,\mu_B^b,\mu_3^b),
\label{GIBBS2}\end{equation}
while for every additive thermodynamical function one has
\[ n_B=\lambda\,n_B^a+(1-\lambda)\,n_B^b,\; {\cal S}=\lambda\,{\cal S}^a+(1-\lambda)\,{\cal S}^b,\;
{\cal E}=\lambda\,{\cal E}^a+(1-\lambda)\,{\cal E}^b, \;
\text{etc.}\] $0\leq \lambda \leq 1$. To evaluate the derivatives of
the thermodynamical potential correctly, the dependence of the
parameter $\lambda(T,\mu)$ must be taken into account.\\
Due to the fact that the thermodynamical potential
$\omega(T,\mu_B,\mu_3)=-P$ is non monotonous, the states $a$ and
$b$ have common values of temperature and chemical potentials and
yet different mean field values  $s, z$ and flavor compositions.
It must be stressed that for $x=0$ there is only one conserved
charge, since in such case the flavors $u, \, d$ are
indistinguishable. Therefore the Gibbs construction does not
apply. For other global asymmetries $x$, pairs of states with
different isospin compositions $x_a,\; x_b$ are associated under
the constraint
\begin{equation}
x=\frac{\lambda\,x_a\,n_a+(1-\lambda)\,x_b\,n_b}{\lambda\,n_a+(1-\lambda)\, n_b}. \label{x Binodal}
\end{equation}
For a fixed temperature these states form a continuum with extreme
values $x_L$ and $x_H$. For a given pressure the collection of all
these points, corresponding to the full range of temperatures
allowed, constitutes the Equilibrium Coexistence Region (CER)
region. To illustrate this subject the ECR corresponding to
$P=43.5$ MeV fm$^{-3}$ is shown in Fig. \ref{Fig Binodal}. It
extends up to a maximum temperature $T=9.2$ MeV, where only states
with $x_a \simeq x_b \simeq 0$ are involved. As a general trend it
is found that the extension of this region reduces when
\textit{i}) the pressure increases, or \textit{ii}) $g$ decreases,
until it completely disappears.

\begin{figure}
    \centering
    \includegraphics[width=0.8\textwidth]{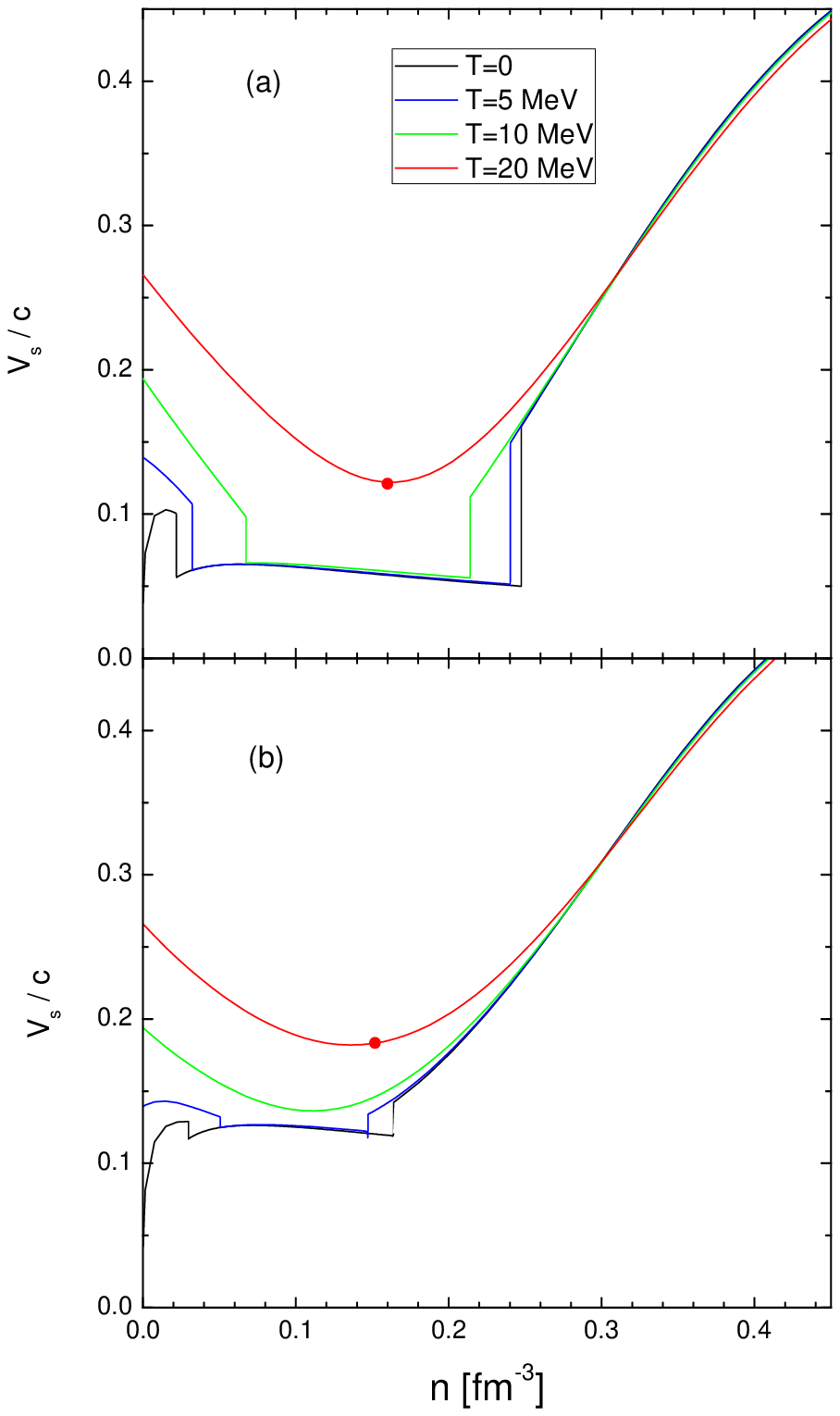}
    \caption{The speed of sound $v_S$ as function of the quark number density for several
    temperatures and $x=1/3$ (a) or $x=2/3$ (b). The reference density is $n_0=0.15$ fm$^{-3}$.
    The vertical segments at the discontinuities have the sole purpose of facilitating the interpretation of the curves.}
    \label{Fig Speed}%
\end{figure}

An interesting quantity which manifests some features of the
equation of state is the speed of sound $v_S$. It has been
proposed that the value of $v_S$ can be inferred from experimental
data of ultrarelativistic heavy ion collisions \cite{GARDIM,MU5}.
It also enters in the definition of the tidal deformability of
binary star systems and non-radial oscillations of compact stars.
For this reason it can indirectly be estimated by the analysis of
the gravitational waves, whose detection is under permanent
attention due to the evolution of methods and technologies. So,
there is a strong motivation for the recent studies of the speed
of sound in quark matter using effective
models \cite{AYALA,AYALA0,GHOSH,HE,HE1,LIU,DOSSOW}. Much of them
explore the limit of $\mu_B \rightarrow 0$ in order to contrast
with LQCD results, but in \cite{DOSSOW} the relation between the
speed of sound and the CEP is studied.\\
In Fig.\ref{Fig Speed} the isothermal speed of sound is shown for
a range  of quark densities and temperatures corresponding to the
first-order transition and two flavor compositions $x=1/3\,(2/3)$
in the upper (lower) panel. For $x=1/3$ and $T=20$ MeV the
temperature is above the CEP and $v_S$ has a non-monotonic trend
with its minimum located approximately at the transition point
(filled circle). For the lower temperatures the system experiences
the coexistence of phases, which reflects in a discontinuous
behavior with a sudden drop of $v_S$. Along this passage the speed
of sound has a mild variation with the particle number, and it is
approximately independent of the temperature. Furthermore, the
shrinking of the ECR region as $T$ grows is evident. A similar
description applies to $x=2/3$, but in such case only the
isotherms corresponding to $T=0, \text{and}\; 5$ MeV undergo a
first order transition.

\begin{figure}
    \centering
    \includegraphics[width=0.8\textwidth]{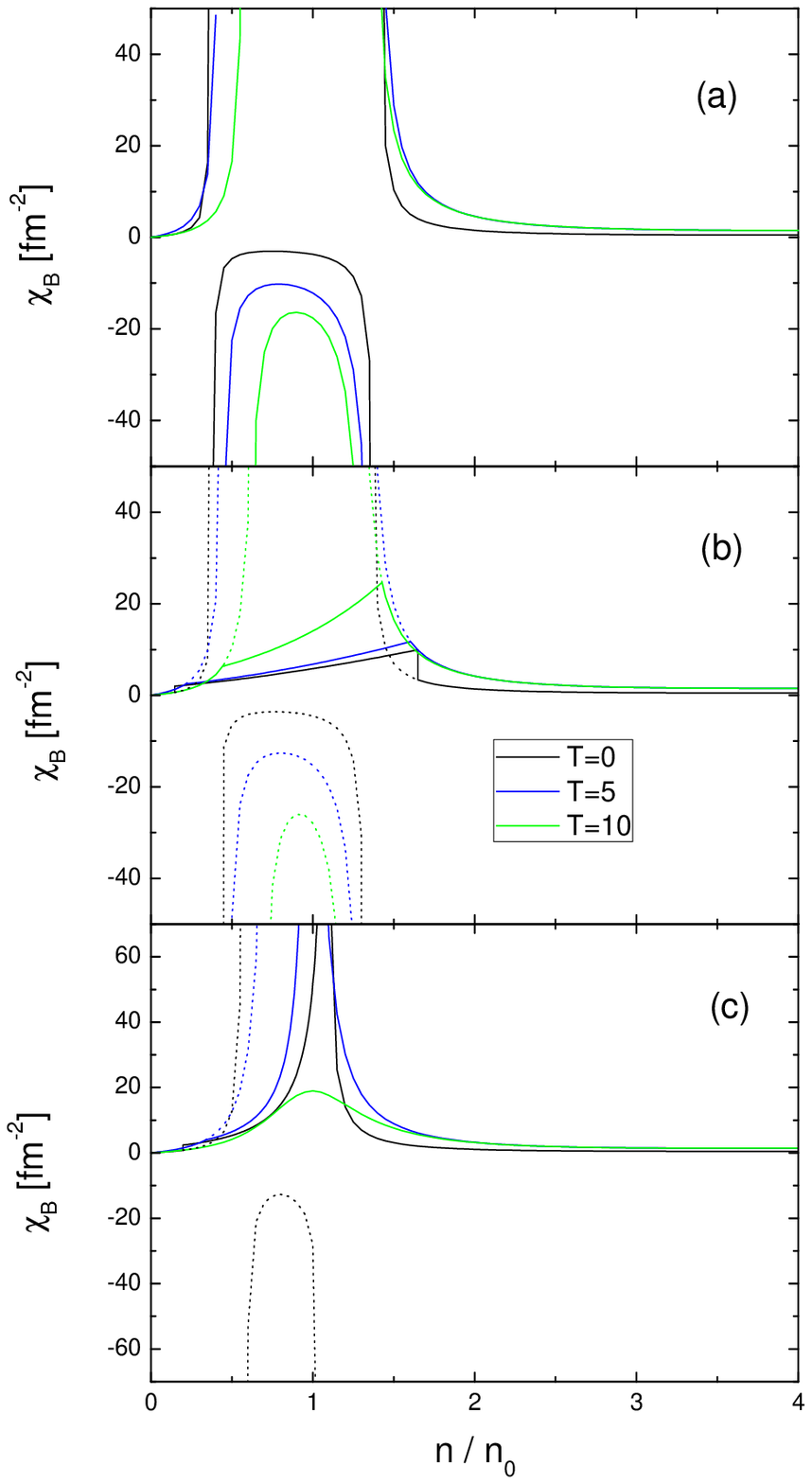}
    \caption{The second order susceptibility corresponding to the baryonic number as
    function of the quark number density for several temperatures and $x=0$ (a), $x=!/3$
    (b), $x=2/3$ (c). For $x=0$ only the results using the unstable equation of state are
    shown. The reference density is $n_0=0.15$ fm$^{-3}$.}
    \label{Fig SuxB}%
\end{figure}

To end this section, the results for the susceptibilities
associated with the conserved charges will be presented. The
effects of the chiral transition on the number susceptibilities
have been discussed previously within the NJL model
\cite{SASAKI,SASAKI1,SASAKI2}, and also with the LSMq
\cite{STOKIC,SKOKOV,SKOKOV1,SKOKOV2,ALMASI}, and particularly the presence of a
spinodal out of thermodynamical equilibrium has been considered \cite{SASAKI1,SASAKI2}.\\
Along the last figures the susceptibilities are shown as functions
of the quark density  for several temperatures in the range of the
first order transition. For  $\chi_B$ (Fig.\ref{Fig SuxB}) strong
oscillations are found, concentrated at the borders of the
binodal. For higher densities all the curves decay and reach the
asymptotic regime quickly. It must be pointed out that within a
Maxwell construction this dramatic behavior is lost since most of
this range of densities is not accessible for the system.\\
For $x=0$ (Fig.\ref{Fig SuxB}a) results including the unstable
region of the equation of state are shown, the peaks and drops are
truly divergences of the same type as those  reported and
discussed in
\cite{SASAKI1,SASAKI2} for calculations using the NJL model.\\
For $x>0$ the figures are more complex because, in contrast to
$x=0$, the system experiences a coexistence of phases. This effect
is clear for $x=1/3$ (Fig.\ref{Fig SuxB}b) where the steep peaks
and drops (dashed lines) are replaced by smooth curves with finite
discontinuities at the end points. In between $\chi_B$ remains
positive and increasing with the density, a fact that is
emphasized with growing $T$.
\\
For the higher $x=2/3$ a qualitative change happens. For $T=10$
MeV the system undergoes a continuous crossover and $\chi_B$ shows
a regular behavior. The cases $T=0, \text{and} \; 5$ MeV, show
noticeable changes since even within the ECR $\chi_B$ has large
but finite discontinuities. However it must be bear in mind that
this is an extrapolation, since the effect of the weak forces will
not be negligible for $x=2/3$.

The susceptibility associated with isospin is completely regular
for $x=0$ (Fig.5a) with an evident enhancement for non-zero
temperatures. For $x=1/3$ (Fig.5b) the unstable region is  marked
by peaks and drops in $\chi_3$, which are replaced by a monotonous
trend within the ECR. Finally, for $x=2/3$ (Fig.5c) the
significant deviations from the null behavior only occurs at the
high-density limit of the binodal and are distinguished by a
divergence for $T=5$ MeV. However the same remarks as for
$\chi_B$, $x=2/3$ apply here.

\section{SUMMARY AND CONCLUSIONS}\label{SUMMARY}

The behavior of the chiral phase transition under asymmetric
flavor  composition of the quark matter has received great
attention in recent years. One of the causes is that LQCD
calculations  have been available, and this is an opportunity for
effective models to improve their schemes by contrasting
predictions with well established results. Since LQCD is efficient
in the limit of zero matter density, much of the recent
theoretical efforts have been restricted to such regime
\cite{SON,STIELE,AVANCINI,LOPES,ADHIKARI,AYALA,AYALA0}. However,
interesting effects have been pointed out in regard of the chiral
transition at non-zero baryonic density. In particular the effects
of fluctuations near the non-equilibrium region of the first order
phase transition have been studied
\cite{SASAKI,SASAKI1,SASAKI2,STOKIC,SKOKOV,SKOKOV1,SKOKOV2,ALMASI}.
By definition this domain can not be analyzed by the equilibrium
thermodynamics. However, if multiple charges are conserved and a
low surface tension  is assumed, the system could undergo a
continuous transition through the coexistence of equilibrium
states \cite{GLENDENNING,HEISELBERG}. These hypothesis have been
developed in this paper in the context of the LSMq with two
flavors. The scalar isoscalar meson $\zeta$  has been included as
an emergent of the isospin imbalance at finite density. The one
loop approximation has been proposed, but calculations have been
finally performed in the MFA. The effects of higher order
corrections is object of present calculations.  Neither pion
condensation nor quark superconductivity have been considered
since the chemical potentials $\mu_B,\; \mu_3$ are below the
corresponding thresholds for the range of this calculations.
\\
The phase diagram obtained is in qualitative agreement with
present knowledge, a  smooth crossover at high temperatures ends
at a CEP and is replaced by a first order transition at low
temperatures. Although the value obtained for the critical
temperature is too low, it is expected that higher order
corrections will improve this result \cite{PETROPOULOS,AYALA1}.
\\
Several important effects on the bulk properties of quark matter
with isospin imbalance have been found and discussed. The speed of
sound at finite isospin and temperatures above the  CEP, have a
non-monotonic dependence on the particle density exhibiting a
minimum corresponding to the crossover transition. Within the
coexistence region $v_S$  experiences a noticeable decrease with
finite discontinuities at the borders. A mild variation with
density and temperature is predicted for this domain.
\\
The susceptibility associated with the baryon number $\chi_B$,
remains bounded for low asymmetry parameter $x>0$, in contrast to
calculations using the unstable sector of the equation of state.
As $x$ grows large but finite discontinuities appear at the high
density border of ECR, but the inclusion of the electromagnetic
field could have important effects in such situations. A similar
comment applies to the susceptibility $\chi_3$, corresponding to
the isospin asymmetry.

\section*{Acknowledgements}
This work has been partially supported by CONICET, Argentina under
the project 11220200102081CO, and by UNLP, Argentina.

\bibliography{ArxivB}
\newpage
\newpage

\newpage
\newpage

\end{document}